\documentclass[twocolumn]{aastex62}
\usepackage{amsmath}
\usepackage{graphicx}
\usepackage{chngcntr}
\usepackage{appendix}
\usepackage{natbib}
\newcommand{\Htwo}{$\rm{H}_{2}$}
\newcommand{\degrees}{$^{\circ}$}
\newcommand{\Msun}{$M_{\odot}$}
\newcommand{\appro}{$\sim$}
\newcommand{\tauv}{$\tau \left(v\right)$}
\newcommand{\tauo}{${\tau }_{{0}}$}
\newcommand{\tauop}{${\tau }_{{0}}^{\prime}$}
\newcommand{\tauf}{${\tau }_{{f}}$}
\newcommand{\tauHI}{${\tau }_{{HI}}$}
\newcommand{\taumax}{${\tau }_{{max}}$}

\newcommand{\TAv}{${T}_{{\rm A}}\left(v\right)$}
\newcommand{\Tbv}{${T}_{{\rm b}}\left(v\right)$}
\newcommand{\Tb}{${T}_{{\rm b}}$}
\newcommand{\Tc}{${T}_{{\rm c}}$}
\newcommand{\ToH}{${T}_{{\rm H}}$}
\newcommand{\Tk}{${T}_{{\rm k}}$}
\newcommand{\Tex}{${T}_{{\rm ex}}$}
\newcommand{\TBo}{${T}_{{{\rm B}_{{0}}}}$}
\newcommand{\To}{${T}_{{1-0}}$}
\newcommand{\Tbg}{${T}_{{\rm bg}}$}
\newcommand{\THINSA}{${T}_{{\rm HINSA}}\left(v\right)$}
\newcommand{\THINSAnonv}{${T}_{{\rm HINSA}}$}
\newcommand{\vH}{${v}_{{\rm H}}$}
\newcommand{\sigmaH}{${\sigma}_{{H}}$}
\newcommand{\sigmaHINSA}{${\sigma}_{{\rm HINSA}}$}
\newcommand{\vel}{$v$}

\newcommand{\sigth}{${\sigma }_{{th}}$}
\newcommand{\sigHth}{${\sigma }_{H_{th}}$}
\newcommand{\sigCOth}{${\sigma }_{CO_{th}}$}
\newcommand{\COIoO}{$^{12}$CO $J=1-0$}
\newcommand{\NHtwo}{$N\left({{H}_{{2}}}\right)$}
\newcommand{\NHINSA}{$N\left({\rm HINSA}\right)$}

\newcommand{\e}[1]{\ensuremath{\times 10^{#1}}}
\newcommand{\XIICO}{$^{12}$CO}
\newcommand{\XIIICO}{$^{13}$CO}

\accepted{by Astrophysical Journal}

\shorttitle{A HINSA Survey of the LMC}
\shortauthors{Liu et al.}

\begin{document}

\title{Tracing the Formation of Molecular Clouds in a Low-Metallicity Galaxy:\\
A HI Narrow Self-Absorption Survey of the Large Magellanic Cloud}

\author{Boyang Liu}
\affil{National Astronomical Observatories, Chinese Academy of Sciences, Beijing, P.R. China, 100012}
\affil{University of Chinese Academy of Sciences, No.19(A) Yuquan Road, Shijingshan District, Beijing, P.R.China 100049}
\affil{CAS Key Laboratory of FAST, National Astronomical Observatories, Chinese Academy of Sciences, Beijing, P.R. China, 100012}
\affil{International Centre for Radio Astronomy Research (ICRAR), University of Western Australia, 35 Stirling Hwy, Crawley, WA 6009, Australia}
\email{liuboyang@nao.cas.cn}

\author{Di Li}
\affil{National Astronomical Observatories, Chinese Academy of Sciences, Beijing, P.R. China, 100012}
\affil{University of Chinese Academy of Sciences, No.19(A) Yuquan Road, Shijingshan District, Beijing, P.R.China 100049}
\affil{CAS Key Laboratory of FAST, National Astronomical Observatories, Chinese Academy of Sciences, Beijing, P.R. China, 100012}
\email{dili@nao.cas.cn}

\author{Lister Staveley-Smith}
\affil{International Centre for Radio Astronomy Research (ICRAR), University of Western Australia, 35 Stirling Hwy, Crawley, WA 6009, Australia}
\affil{ARC Centre of Excellence in All Sky Astrophysics in 3 Dimensions (ASTRO 3D)}
\email{lister.staveley-smith@uwa.edu.au}

\author{Lei Qian}
\affil{National Astronomical Observatories, Chinese Academy of Sciences, Beijing, P.R. China, 100012}
\affil{CAS Key Laboratory of FAST, National Astronomical Observatories, Chinese Academy of Sciences, Beijing, P.R. China, 100012}

\author{Tony Wong}
\affil{University of Illinois, 227 Astronomy Building, MC-221, 1002 W. Green St., Urbana, IL 61801, USA}

\author{Paul Goldsmith}
\affil{Jet Propulsion Laboratory, M/S 180-703, 4800 Oak Grove Drive, Pasadena, CA 91109}

\begin{abstract}
Cold atomic hydrogen clouds are the precursors of molecular clouds. Due to self-absorption, the opacity of cold atomic hydrogen may be high, and this gas may constitute an important mass component of the interstellar medium (ISM). Atomic hydrogen gas can be cooled to temperatures much lower than found in the cold neutral medium (CNM) through collisions with molecular hydrogen. In this paper, we search for HI Narrow Self-Absorption (HINSA) features in the Large Magellanic Cloud (LMC) as an indicator of such cold HI clouds, and use the results to quantify atomic masses and atomic-to-molecular gas ratio. Our search for HINSA features was conducted towards molecular clouds in the LMC using the ATCA+Parkes HI survey and the MAGMA CO survey. HINSA features are prevalent in the surveyed sightlines. This is the first detection of HINSA in an external galaxy. The HINSA-HI/\Htwo\ ratio in the LMC varies from 0.5\e{-3} to 3.4\e{-3} (68\% interval), with a mean value of $(1.31 \pm 0.03)$\e{-3}, after correcting for the effect of foreground HI gas. This is similar to the Milky Way value and indicates that similar fractions of cold gas exist in the LMC and the Milky Way, despite their differing metallicities, dust content and radiation fields. The low ratio also confirms that, as with the Milky Way, the formation timescale of molecular clouds is short. The ratio shows no radial gradient, unlike the case for stellar metallicity. No correlation is found between our results and those from previous HI absorption studies of the LMC.
\end{abstract}


\section{Introduction}

It is generally accepted that molecular clouds are the birth places of stars \citep[and references therein]{MO2007}. In the classic scenario \citep{Shu1987}, pre-star-forming molecular clouds are spherically layered structures with the molecular, atomic and ionized gas phases assumed to be dominant from the inside to the outside. Various molecular tracers have been used to trace \Htwo, eg. [C I] \citep{Pin2017,Val2018,Oka2019}, [C II] \citep{Tang2016,Zan2018,Ryb2019}, OH \citep{DWM2019,EA2019,Tang2017}, the main content of molecular clouds, with CO being one of the most widely used \citep[e.g.][]{HD2015,Gen2015,MWISP2019}. By assuming a fixed dust-to-gas ratio, FIR and millimeter continuum observations can be used to indicate the total gas content, including atomic and molecular components \citep[e.g.][]{Bot2007,GR2014,Lenz2017}, although there could be a bias from inaccurate assumption of the dust-to-gas ratio. The 21cm line is generally used to trace the atomic hydrogen (HI) component and is considered to be optically thin in most situations \citep{GH1988}. Recombination lines as well as centimeter continuum are often used to trace the ionized gas component \citep{Ind2009}.

While these tracers are able to depict the general picture of the different phases of gas, they have obvious weakness. CO and other molecules only trace \Htwo\ above certain densities and extinctions, and their abundance can be easily biased by local metallicity \citep{Mad2016}. The excitation temperature and optical depth cannot be simultaneously determined from a single line, so the column density can easily be underestimated, even for the HI 21cm line \citep{Ber2008,Hei2003b,Dic2003,ST2004,Dic2009}.

Deriving the total amount of cold HI gas by analyzing self-absorption features of the 21cm line is feasible, but is complicated by confusion due to multiple components \citep{RC1972,Gib2000,Kav2003,MCG2006,Den2018}. More sophisticated approaches to the analysis of HI self-absorption have been made during the past decade: \citet{LG2003} proposed the concept of HI Narrow Self-Absorption (HINSA) to refer to the HI self-absorption features associated with cold HI gas mixed in molecular cores, following the discovery of narrow HI absorption features coinciding with OH emission lines in a number of Galactic clouds. These authors derived the column density of cold HI gas indicated by HINSA features. By constructing a time-dependent molecular cloud formation model in which the rate of transformation of HI to \Htwo\ by dust surface chemistry balances the \Htwo\ destruction rate due to cosmic rays, \citet{GL2005} utilized the cold-HI/\Htwo\ ratio using HINSA features as a chemical clock to probe the formation of molecular clouds. This proved the HINSA technique as a new tool to study the early state of molecular cloud formation. \citet{Tian2010} have also shown that HINSA technique can be adopted as an indicator of spatial relationship between features.

\citet{LG2003} reported a HINSA detection rate of 77\% for the clouds in Taurus/Perseus region. \citet{Krco2010} found a detection rate of over 80\% over a wide range of environments in the Galaxy. The prevalence of HINSA features suggest cold HI gas is always associated with molecular cores, at least in our Galaxy. It is therefore of interest to explore a different environment to test for the presence of HINSA features, and study the properties and evolution of molecular clouds using this technique.

The Large Magellanic Cloud (LMC) is an ideal target for a similar study. As the nearest gas-rich galaxy to the Milky Way, it is located at a distance of 50kpc \citep{W1997,Pie2013,deG2014}. Its prominent disk has a low inclination angle of 33\degrees\ \citep{W1997}, i.e.\ it is close to face-on. This permits spatially resolved studies of the galaxy's stellar and ISM content, making the study of the LMC more similar to ``galactic" than ``extragalactic" environments. With a smaller stellar mass of a few $10^9$ \Msun\ \citep{Fei1980,Kim1998,AN2000}, the LMC is in a more primitive evolutionary state than the Milky Way and other large disk galaxies: its ISM metallicity is 0.2 dex lower than the local value \citep{RD1992,W1997,RD2019}, consistent with the trend of lower-mass galaxies having lower metallicity \citep[e.g.][]{Tre2004,Kew2008,Asa2009,Man2010,Sch2015}. Thus studies of the LMC have the potential to reveal the `gastrophysics' (gas astrophysics) and star formation laws of similar low-metallicity irregular galaxies in the high-redshift Universe \citep{Wil2009}.

Several studies of the cool phase HI in the LMC have been conducted in the past two decades. \citet{Dic1994} and \citet{Dic1995} suggested that the cool gas in the LMC is either more abundant or colder than that of the Milky Way by analyzing the absorption spectrum of background compact continuum sources. \citet{Meb1997} and \citet{MZ2000} confirmed this trend and reported typical temperatures of the diffuse cool gas in the LMC of 30-40 K, compared with the typical value of 60 K in the solar neighborhood \citep{Kal1985}. \citet{Bra2012} used a different approach of Gaussian component fitting and found a low temperature for the LMC cool gas consistent with previous studies. This study also created an opacity-corrected HI column density map of the LMC, finding a global correction factor of 1.33. Infrared \citep{Ber2008,Gal2011,Meix2013} and ultraviolet \citep{Tum2002,Wel2012,RD2019} studies have also provided important information on cool phase atomic gas in the LMC.

Different techniques have been applied in previous HI absorption studies of the LMC. However the HINSA technique has never been utilized beyond the Milky Way. With the advent of a recent LMC CO survey, i.e. the MAGMA survey \citep{Wong2011} using the ATNF 22 m Mopra telescope, it is now possible to probe cold HI gas associated with molecular cores using the HINSA technique applied to the MAGMA CO cloud catalog. We have therefore conducted a joint analysis of the MAGMA CO data cube and the ATCA+Parkes HI survey data \citep{Kim2003} to study the properties of the HINSA cold HI gas in the LMC.

Section 2 of the paper describes the data; Section 3 explains the data reduction process using different HINSA techniques; Section 4 shows the main results and Section 5 discusses the applicability of different HI absorption techniques and the implications for the LMC. Finally, we summarize our result in Section 6.

\section{Data}
In this section we introduce the data used in this study. 

\begin{figure*}
\centering
\includegraphics[width=7in]{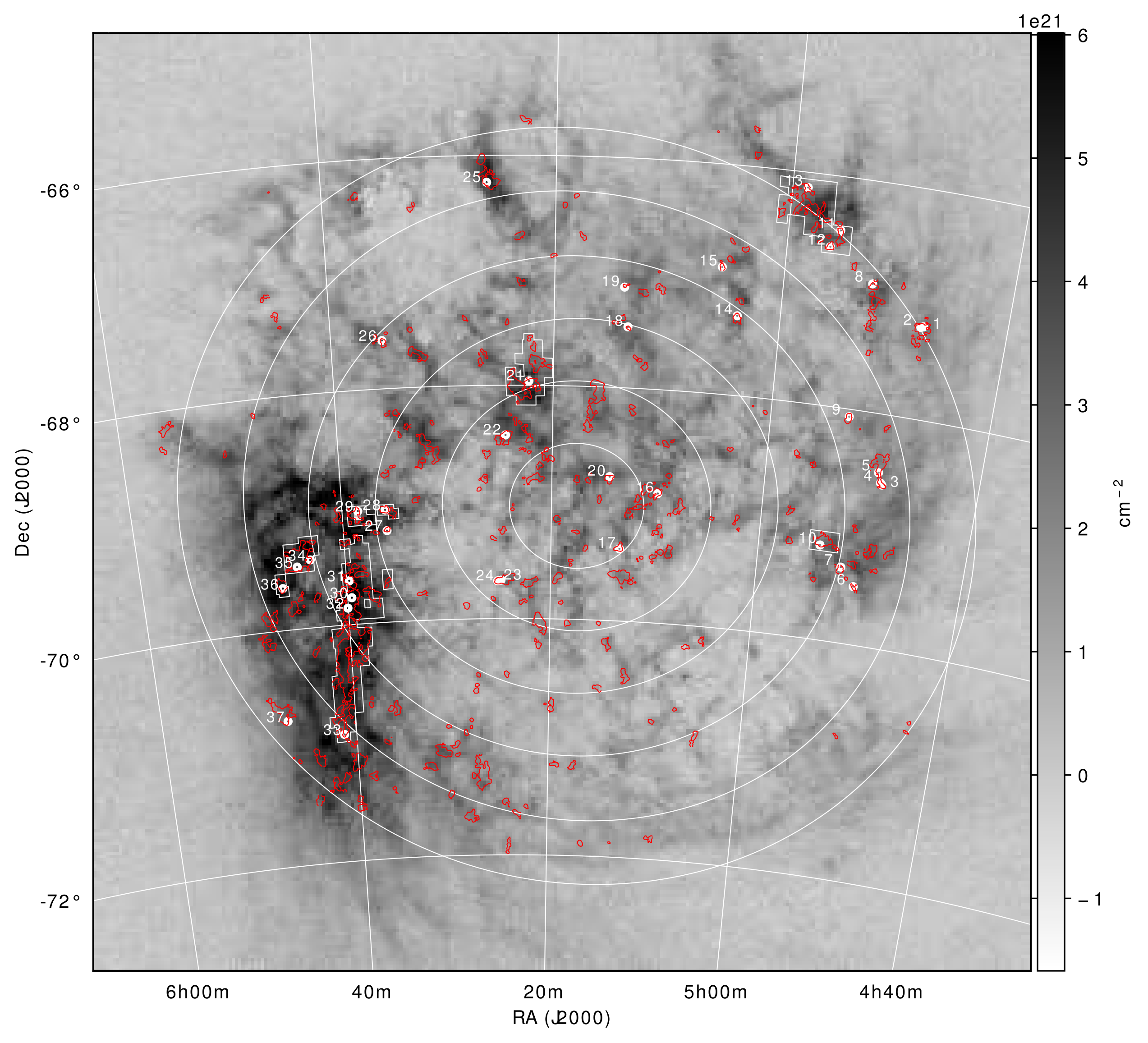}
\caption{Data used in this work. Grayscale image: HI column density from the ATCA+Parkes LMC HI Survey \citep{Kim2003}; red contours: the MAGMA CO Survey DR3 moment 0 map, with a contour level at 1.0 K$\cdot$km/s; white rectangular regions: boundaries of the MAGMA \XIIICO\ map for selected regions; white ellipses: radial rings as described in Section 4.3; white circle markers and green labels: the location and ID of the sources listed in Table 1.}
\label{fig:data_map}
\end{figure*}

\subsection{HI}
An HI 21cm survey with resolution of 1\arcmin.0 (\appro15 pc assuming a distance of 50 kpc) was conducted during the late 1990s with the Australia Telescope Compact Array (ATCA)\citep{Kim1998}. Due to the missing flux problem for interferometers, this survey was not sensitive to structures larger than 500 pc. To complement these data,  \citet{Kim2003} combined ATCA interferometer and Parkes single-dish observations \citep{SS2003} to give the most complete HI survey of the LMC in terms of sky and spatial frequency coverage. Their data cube contains a complete sampling of spatial structures from 15 pc to 10 kpc. The velocity resolution is 1.649 km\,s$^{-1}$ and brightness temperature sensitivity 2.4 K.

\subsection{CO}
The most complete CO survey in terms of sky coverage in the past decade has been the second LMC CO survey conducted by the NANTEN telescope \citep{Fuk2008}. It is a spatially continuous survey which identified 272 molecular clouds. The Magellanic Mopra Assessment (MAGMA) is a follow-up CO survey to target detected regions, with better sensitivity by a factor of 2, and was conducted with the ATNF 22m Mopra telescope \citep{Hug2010}. \citet{Wong2011} cataloged 450 molecular clouds based on the CO $J$=1-0 map.

We employ the third data release of MAGMA for this study \citep{Wong2011,Wong2017}. It contains the CO $J$=1-0 cube described in \citet{Wong2011}. The cube has an angular resolution of 45\arcsec, and a pixel spacing of 15\arcsec. The velocity resolution is 0.526 km\,s$^{-1}$. The rms noise of the cube is typically 300 mK. Compared to the published paper \citep{Wong2011}, the released data cube has been processed with a constant 10 mK offset to bring the baseline back to \appro 0 K.

As described in Sections 4.1 and 5.1, we also utilized the unreleased MAGMA \XIIICO\ data for optical depth determination.
\XIIICO\ observations were obtained simultaneously with the \XIICO\ observations for data obtained in 2006 June to 2013 September, and will be described fully in a separate paper (Wong et al., in preparation).
A merged cube was generated from 1244 individual 5\arcmin\ $\times$ 5\arcmin\ square maps spanning a heliocentric velocity range of 200--325 km s$^{-1}$.  
The CO spectra were placed on a main-beam brightness temperature scale ($T_{\rm mb}$) assuming an ``extended beam'' efficiency of 0.43 based on daily observations of Orion KL referenced to the measurements of \citet{Ladd2005}.  
Our $T_{\rm mb}$ scale has recently been confirmed by comparison with ALMA total power mapping (R. Indebetouw, private communication).
The resulting maps possess a Gaussian beam of 45\arcsec\ FWHM which is oversampled with a pixel scale of 15\arcsec.  
The typical RMS map noise is $\sigma(T_{\rm mb}) \approx 0.19$ K per 0.55 km s$^{-1}$ channel.

The spatial coverage of the CO and \XIIICO\ data used in this study is shown in Figure~\ref{fig:data_map}, on top of the HI column density map for the LMC.

\section{Methods}

\subsection{HINSA techniques}
One challenge to applying the HINSA concept to analysis of HI absorption features is how to reconstruct the background emission or the ``original" spectrum before absorption. An accurately recovered ``original" spectrum leads to an accurately defined absorption line profile, and vice versa. Previous studies have used several different approaches.

\citet{LG2003} adopted an intuitive method by masking the absorption feature and fitting the rest of the HI profile with a polynomial. This is common practice in absorption analysis, but suffers from the subjectivity in judging the shape of the original spectrum. As they reported, the fitted result can vary as much as 1 K using different orders of polynomial. \citet{Per2011} made the assumption of a smooth and gradual variation of the background emission, and take the average spectrum of several reference points around the center of the core as the ``original'' spectrum. But as stated by many authors, the HI gas is intrinsically filamentary \citep[e.g.][]{Elm2011}, thus considering it as ``smooth and gradual'' can cause unpredictable biases.

\citet{Krc2008} presented a new technique to improve the quality of HINSA feature fitting procedure. Considering the narrow nature of HINSA features, they proposed that the narrow dip in the HI profile would generate a feature in the 2nd-derivative of the observed line profile since the slowly changing ``original'' profile is largely suppressed while the fast changing absorption dip is highlighted. This was used to locate the HINSA-like absorption features in the HI profile. By constraining the regions searched by such a method with molecular tracers, finding the possible HI self-absorption features associated with molecular clouds is possible. This provides a more convenient way to extract the HINSA profile with more confidence than the previous methods.

\subsection{HINSA techniques applied in this work}
In this work, we basically adopt the \citet{Krc2008} technique, although some modifications were made to cope with the fact that the MAGMA program had only released \XIICO\ data at the time of our analysis. 

\subsubsection{Radiative transfer analysis}

Assuming the cold HI gas responsible for a HINSA feature has optical depth \tauv, then:
\begin{equation}
{T}_{{A}}\left({v}\right)={T}_{{b}}\left({v}\right){e }^{{-\tau \left({v}\right)}}+{T}_{{H}}\left[{1-{e }^{{-\tau \left({v}\right)}}}\right],
\end{equation}
where \vel\ is the velocity, \TAv\ is the observed HI spectrum, \Tbv\ is the background HI emission or so-called ``original'' spectrum,  including the emission from background HI clouds as well as other background sources such as the CMB. \ToH\ is the temperature of the HINSA-generating cold HI associated with molecular material. In writing this function, we have neglected the foreground warm HI which is actually not affected by the absorbing cold HI gas. The same approximation was adopted by \citet{Krc2008} for the nearby sources in the Galaxy. For the sources in the LMC that could be embedded anywhere in the HI disk,  this could be a poorer assumption. The impact of this will be discussed later.

We make the simple assumption that \tauv\ has a Gaussian shape and can be expressed as
\begin{equation}
\tau \left({v}\right)={\tau }_{{0}}\exp\left({-\frac {{{\left({v-{v}_{{H}}}\right)}}^{{2}}}{2{\sigma }_{{H}}^{{2}}}}\right),
\end{equation}
where \tauo\ represents the peak optical depth of the cold HI gas, \vH\ is the velocity of the peak optical depth, and \sigmaH\ is the width of the optical depth profile. In our study, we use a single Gaussian fit to the CO spectrum, and take the fitted central velocity of the CO peak as the value of \vH. The line width of the gas component \sigmaH, consists of two components, thermal and non-thermal according to:
\begin{equation}
	{\sigma}_{{H}} ={{\left({{\sigma }_{{H_{th}}}^{{2}}+{\sigma }_{{H_{nt}}}^{{2}}}\right)}}^{{\frac {1}{2}}},
\end{equation}
where the subscripts \textit{th}\ and \textit{nt}\ represent thermal and non-thermal, respectively. Similarly, for the CO gas:
\begin{equation}
	{\sigma}_{{CO}} ={{\left({{\sigma }_{{CO_{th}}}^{{2}}+{\sigma }_{{CO_{nt}}}^{{2}}}\right)}}^{{\frac {1}{2}}}.
\end{equation}

For well-mixed gas, the non-thermal line width would be similar for different components \citep{LG2003}. Combining formulas (3) and (4), we obtain:

\begin{equation}
{\sigma}_{{H}} =\left[{{\sigma}_{{CO}}^{{2}}+{\left({{\sigma }_{{H_{th}}}^{{2}}-{\sigma }_{{CO_{th}}}^{{2}}}\right)}}\right]^{{\frac {1}{2}}},
\end{equation}
where the thermal linewidth for both HI and CO gas satisfy

\begin{equation}
	{\sigma }_{{th}}={{\left({\frac {2kT}{m}}\right)}}^{{\frac {1}{2}}},
\end{equation}
where $m$ represents the mass of a hydrogen atom or CO molecule, when \sigth\  is replaced by \sigHth\  or \sigCOth, respectively. Assuming that the different gas components inside the molecular cloud are in thermodynamic equilibrium then, for either HI or CO, the temperature $T$ in equation (6) can be replaced with the same CO kinetic temperature \Tk. Under the assumption of LTE, we take \Tk\ to be equal to \Tex, the excitation temperature of CO. We therefore have

\begin{equation}
	f\left({{T}_{{ex}}}\right)=\frac {{T}_{{{B}_{{0}}}}}{{T}_{{1-0}}}+f\left({{T}_{{bg}}}\right),
\end{equation}
where $f\left(T\right)$ is defined as

\begin{equation}
	f\left({T}\right)=\frac {1}{{\exp{\left({\frac {{T}_{{1-0}}}{T}}\right)}}-1}.
\end{equation}
\TBo\ is the brightness temperature at the CO line center, here adopted as the peak temperature of the fitted Gaussian profile. \To\ is the equivalent temperature of the \COIoO\ transition and has the value 5.53~K. \Tbg\ is the background field temperature, for which we use the CMB temperature of 2.73 K.

With these assumptions and relations, we can recover the ``original'' HI spectrum as function of a single variable \tauo. 

As demonstrated in \citet{Krc2008}, a narrow dip in a smooth line would generate a prominent feature in the 2nd-derivative profile. Ideally, we expect that such a feature can be minimized if we adjust the value of \tauo\ until the narrow dip in spectrum vanishes. We integrate the square of the 2nd-derivative of the recovered ``original'' spectrum, and  we stop adjusting the value of \tauo\ when the change falls below a precision criterion of $10^{-4}$ K km s$^{-1}$. We are left with the peak optical depth \tauo\ and the ``original'' HI spectrum before absorption, \Tbv.

The amount of HINSA absorption as a function of frequency, i.e. the HINSA profile, is the difference between the ``original'' HI spectrum and the observed HI spectrum. Then we can derive the HINSA brightness temperature profile as

\begin{equation}
\begin{aligned}
{T}_{{\rm HINSA}}\left({v}\right)&={T}_{{b}}\left({v}\right)-{T}_{{A}}\left({v}\right)\\
&=\left({{T}_{{A}}\left({v}\right)}-{{T}_{{ex}}}\right)\left({e }^{{\tau \left({v}\right)}}-{1}\right).
\end{aligned}
\end{equation}

In summary, we derive the HINSA profile using the following stpdf:
\begin{itemize}
  \item Calculate the so-called ``original'' HI spectrum, which does not show the absorption and thus appears smoother (the smoothness of the ``original'' spectrum is judged by its 2nd-derivative). 
  \item Subtract the real HI spectrum from the calculated ``original'' spectrum to derive the HINSA profile.
\end{itemize}

\subsection{Deriving physical parameters}
It can be seen from Equation 9 that when the central velocities for \tauv\ and the observed HI spectrum \TAv\ are different, an asymmetric \THINSA\ profile will result.
To parameterize the \THINSA\ profile, a Gaussian fit is performed, and peak temperature, central velocity and width \sigmaHINSA\ are derived. 

Then we calculate the column density of the HINSA-associated cold HI based on \citet{LG2003}'s formula (13):

\begin{equation}
\begin{aligned}
\frac{N\left({\rm HINSA}\right)}{{\rm cm}^{-2}}=1.95\times {{10}}^{{18}}{\tau }_{{0}}\frac {{\sigma }_{{\rm HINSA}}}{\rm km\;s^{-1}}\left(\frac {T_k}{\rm K}\right) .
\end{aligned}
\end{equation}

CO is almost always optically thick in molecular cores, so the estimation of \Htwo\ column density \NHtwo\ based solely on CO can be unreliable. However, the CO luminosity-\Htwo\ column density conversion factor, or X-factor, is often the only way to estimate \Htwo\ column density in external galaxies \citep[and references therein]{Bol2013}. Similarly for the LMC, there is currently no other molecular tracer available that has such completeness in coverage. We therefore use the latest estimate for the LMC X-factor, $4\times {{10}}^{{20}}{\rm cm}^{{-2}}\left({\rm K\;km\;s^{-1}}\right)^{-1}$  \citep{Bol2013}, which is a direct result of the MAGMA Project \citep{Hug2010,Wong2011,Pin2010}. 

The integrated flux of the CO profile is calculated from the Gaussian fit to avoid the effect of component blending. The HI-to-H2 ratio, defined as the ratio in the HINSA-associated cold HI and the \Htwo\ content, is calculated by comparing \NHINSA\ and \NHtwo. It is the major parameter we derive that shows the abundance of the HINSA-associated HI.

\subsection{Optical depth correction}
The molecular clouds where HINSA features are detected are embedded in the LMC's HI gas disk. The presence of foreground HI gas diminishes the strength of the HINSA absorption dip that we're looking for. Unlike \citet{LG2003} who estimate the proportion of the foreground gas using the Galactic rotation curve, the location of the molecular clouds within the LMC disk is unknown. Here we evaluate the effect of the foreground gas on the observed HINSA features.

Using the same variable $p$ as \citet{LG2003} to describe the position of a given molecular cloud in a uniform disk, $(1-p)$ is the fraction of foreground HI gas relative to the total amount of HI gas in the line of sight. The \emph{real} optical depth of the HINSA HI (equation 12 of \citet{LG2003}) is:

\begin{equation}
\begin{aligned}
{\tau}_{0}^{\prime}=\ln \left[\frac{{p}{T}_{b}+\left({T}_{c}-{T}_{H}\right)\left(1-{\tau}_{f}\right)}{{p}{T}_{b}+\left({T}_{c}-{T}_{H}\right)\left(1-{\tau}_{f}\right)-{T}_{\rm HINSA}}\right]
\end{aligned}
\end{equation}
where \Tb, \ToH\ and \THINSAnonv\ are as defined in section 3.2.1, \Tc\ is the continuum temperature, \tauf\ is the foreground HI optical depth, and ${\tau }_{{f}}=\left({1}-{p}\right){\tau }_{{HI}}$, where \tauHI\ is the total HI optical depth along the line of sight through the LMC's disk. When the foreground HI is ignored as it was done in section 3.2.1, $p=1$ and \tauop=\tauo\ as defined in section 3.2.1. The optical depth correction factor, given by $C$, is defined by:

\begin{equation}
\begin{aligned}
{C}=\frac{{\tau}_{0}^{\prime}}{{\tau}_{0}}.
\end{aligned}
\end{equation}

\begin{figure}
\centering
\includegraphics[width=3.3in]{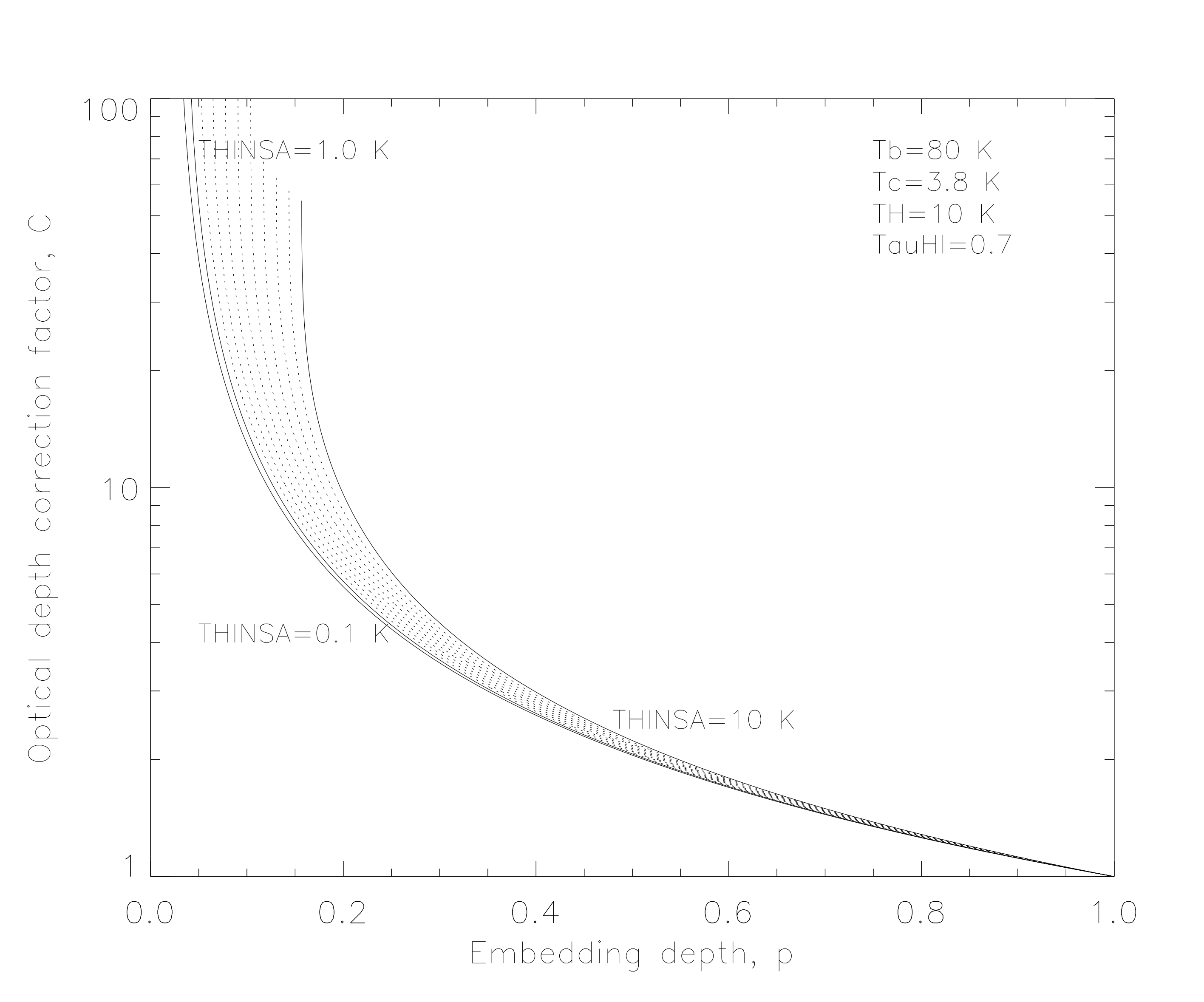}
\caption{The optical depth correction factor $C$ for different values of the embedding depth $p$ and $T_{\rm HINSA}$.}
\label{fig:correctionfactor}
\end{figure}

Using a typical set of parameters \Tb$={80}$ K (Galactic value, \citet{LG2003}), \Tc$={3.8}$ K, \ToH$={10}$ K and \tauHI$=0.7$ , a typical $C(p,\ $\THINSAnonv$)$ relation is shown in Figure~\ref{fig:correctionfactor}. As shown in Figure \ref{fig:A1}, the adopted value \Tb$={80}$ K is also a typical value in the LMC for HINSA regions.

\citet{LG2003} adopted \Tc$={3.5}$ K for Milky Way studies, whereas the value used for the LMC (3.8 K) is derived from the 20-cm continuum map of \citet{Hugh2007}. The flux for Region 3 of \citet{Hugh2007}, where the continuum flux at 3.75 GHz \citep{Hayn1991} is higher than 40 mJy beam$^{-1}$, i.e. the brighter part of the LMC, is considered for the derivation of \Tc. 
The value adopted for \tauHI\ is the average value of \taumax\ measured by \citet{Dic1994}, \citet{Meb1997} and \citet{MZ2000} towards 87 radio sources behind the LMC.

The value of $C$ is large when $p$ is small, and is very sensitive to \THINSAnonv. However for $p>0.3$, the scatter becomes smaller and $C$ approaches unity. Although the exact value of $p$ is unknown, assuming that the scale height of the molecular disk of the LMC is smaller than that of the HI disk, we can adopt $p={0.5}$. For $p\sim0.5$, the value of $C$ varies in a narrow range of $\sim2$ for different values of \THINSAnonv\ -- the difference is less than 14\% for values between 0.1 K and 10 K. In the following calculation, \Tc\ is fixed to ${3.8}$ K, while the values of \Tb, \ToH\ and \THINSAnonv\ are used on a pixel-by-pixel basis.

\section{Results}

Of all the $1997\times 2230$ pixels in the HI data cube, 1446 pixels were detected with HINSA features, i.e. an angular filling factor of $\sim 3\times 10^{-4}$. The details of these detections are given below.

\subsection{HINSA-HI abundance}
Figure~\ref{fig:HIH2ratio} shows the histogram of HINSA-HI abundance, i.e.\ the ratio of HINSA-HI column density to that of \Htwo, which is important for comparison with other studies. 

The value after optical depth correction varies from 0.5\e{-3} to 3.4\e{-3} (68\% interval), with a mean value of $(1.31 \pm 0.03)$\e{-3}; the value before correction varies from 0.3\e{-3} to 1.6\e{-3} (68\% interval), with a mean value of $(0.64 \pm 0.02)$\e{-3}. We also show the results from \citet{LG2003}, a HINSA survey of the Taurus Molecular Cloud and \citet{Krco2010}, a HINSA survey in other regions in the Milky Way for comparison. The 68\% interval value range is 0.2\e{-3} to 4.4\e{-3} for \cite{LG2003}, 0.5\e{-3} to 2.5\e{-3} for \citet{Krco2010}, and 0.4\e{-3} to 3.0\e{-3} for both Milky Way samples combined. The mean value for both Milky Way samples is $(1.0 \pm 0.2)$\e{-3}. Our result shows that the LMC's HINSA-HI/\Htwo\ abundance ratio is slightly higher, but not significantly different from the Milky Way value, which means the LMC has a similar cold gas fraction to the Milky Way.

\begin{figure*}
\centering
\includegraphics[width=6in]{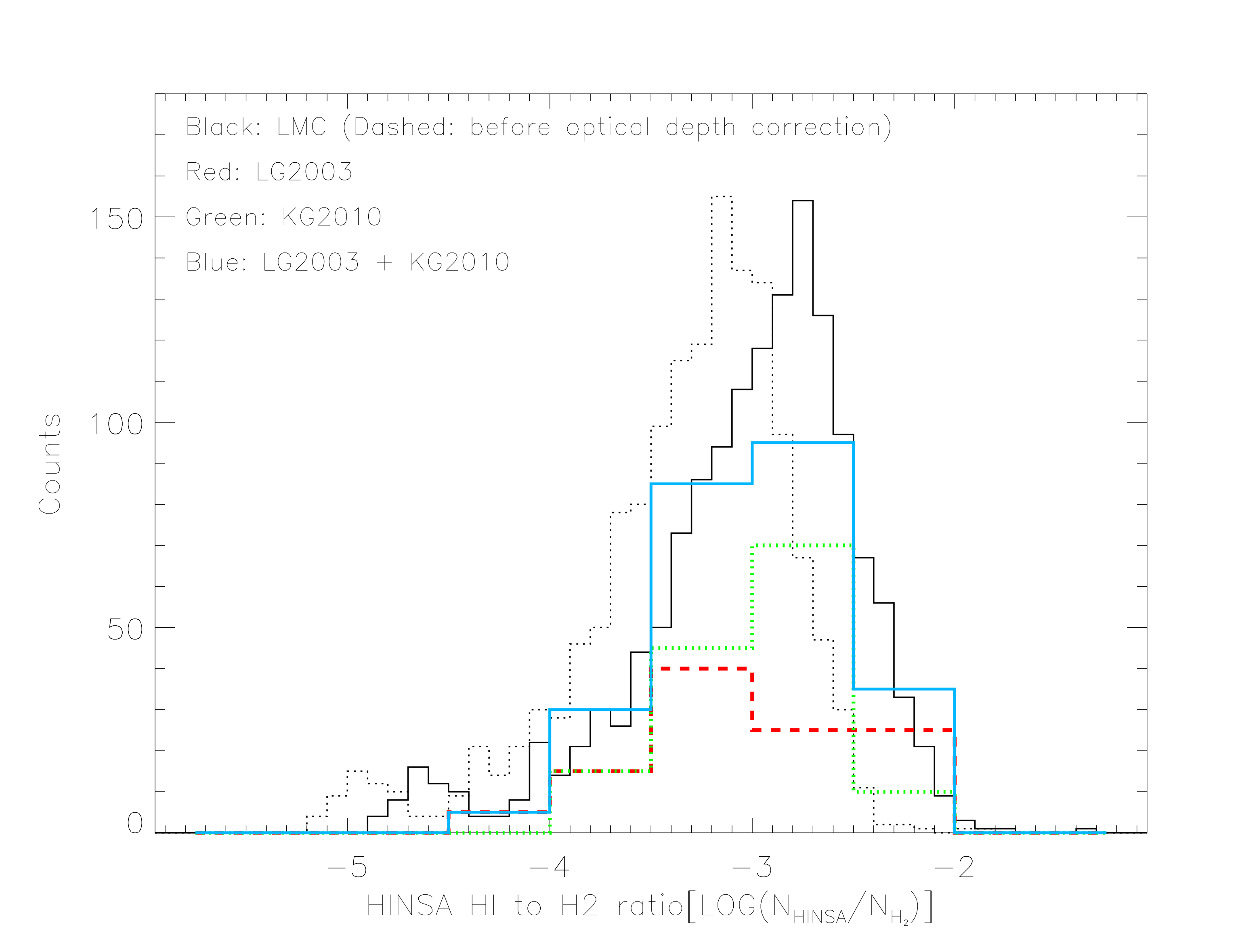}
\caption{A histogram of the HINSA-HI to \Htwo\ ratio $\log_{10} (N_{\rm HINSA}/N_{\rm H_2}$). The bold black histogram is the LMC results from the present work. The red histogram shows the result for Taurus/Perseus region from \citet{LG2003}. The green histogram shows the result for Milky Way regions outside Taurus from \cite{Krco2010} (the values for each velocity component instead of the mean value for each line-of-sight are used). The blue histogram shows the sum of the previous two studies. To improve the visibility of the diagram, the $y$-axis is scaled up by a factor of 5 for the Milky Way results.}
\label{fig:HIH2ratio}
\end{figure*}

\subsection{Catalog}
The HINSA detections were inspected manually. Consecutive pixels with detections were catalogued into the same ``group''. There are 37 groups of HINSA detections in the LMC where the peak optical depth of HINSA-HI is higher than 0.2. Table \ref{tbl:selectedsightlines} is a catalog of the physical parameters of the peak optical depth positions for these groups.

\begin{table*}
\centering
\begin{tabular}{lccccccccr}
\hline
\hline
No. & $\alpha$ (2000) & $\delta$ (2000) & $\tau_{0}$ & $T_{H}$  & $\sigma_{H}$ & $N_{\rm HINSA}$  & $N_{\rm H_2}$     & $N_{\rm HINSA} /N_{\rm H_2}$ & Cloud ID  \\
  & (h:m:s) & (\arcdeg:\arcmin:\arcsec) &            & (K)      & (km\,s$^{-1}$)       & 
  (cm$^{-2}$)  & (cm$^{-2}$) &                &  		  \\\hline
1  & 04:47:21.90 & -67:11:42.3 & 0.29 & 4.0 & 2.0 & 7.6E+18 & 2.5E+21 & 7.0E-03 & 9   \\
2  & 04:47:34.98 & -67:12:16.0 & 0.31 & 4.5 & 1.9 & 8.3E+18 & 2.3E+21 & 8.8E-03 & 10  \\
3  & 04:49:01.79 & -68:36:17.2 & 0.36 & 5.2 & 1.8 & 1.2E+19 & 3.9E+21 & 7.3E-03 & 19  \\
4  & 04:49:11.07 & -68:35:03.9 & 0.30 & 4.9 & 0.9 & 4.0E+18 & 1.6E+21 & 5.8E-03 & 24  \\
5  & 04:49:29.52 & -68:30:14.5 & 0.21 & 4.3 & 1.9 & 5.7E+18 & 2.5E+21 & 5.1E-03 & 30  \\
6  & 04:50:23.76 & -69:30:15.9 & 0.31 & 4.7 & 0.7 & 3.5E+18 & 1.1E+21 & 7.5E-03 & 36  \\
7  & 04:51:50.21 & -69:21:18.0 & 0.31 & 4.3 & 1.4 & 6.2E+18 & 2.4E+21 & 6.0E-03 & 44  \\
8  & 04:52:16.74 & -66:53:40.6 & 0.24 & 4.0 & 1.5 & 4.7E+18 & 1.8E+21 & 6.0E-03 & 50  \\
9  & 04:52:51.04 & -68:03:51.5 & 0.31 & 4.7 & 1.7 & 8.2E+18 & 3.1E+21 & 6.1E-03 & 58  \\
10 & 04:54:05.70 & -69:11:33.1 & 0.29 & 5.1 & 2.6 & 1.3E+19 & 5.0E+21 & 5.9E-03 & 65  \\
11 & 04:55:33.86 & -66:28:16.9 & 0.33 & 4.8 & 2.3 & 1.2E+19 & 4.6E+21 & 6.3E-03 & 78  \\
12 & 04:56:17.62 & -66:37:26.5 & 0.26 & 5.0 & 1.4 & 6.2E+18 & 2.8E+21 & 5.1E-03 & 80  \\
13 & 04:58:42.28 & -66:07:59.2 & 0.20 & 5.3 & 1.7 & 5.9E+18 & 3.8E+21 & 3.5E-03 & 110 \\
14 & 05:03:47.65 & -67:18:35.1 & 0.20 & 5.5 & 1.8 & 6.3E+18 & 4.5E+21 & 3.1E-03 & 137 \\
15 & 05:05:26.14 & -66:53:54.0 & 0.24 & 4.6 & 1.4 & 5.1E+18 & 2.2E+21 & 5.4E-03 & 146 \\
16 & 05:09:55.96 & -68:53:33.3 & 0.22 & 4.5 & 2.7 & 8.7E+18 & 4.1E+21 & 4.7E-03 & 165 \\
17 & 05:13:21.03 & -69:23:03.4 & 0.24 & 7.4 & 1.8 & 9.0E+18 & 7.7E+21 & 2.6E-03 & 207 \\
18 & 05:13:25.50 & -67:28:17.6 & 0.31 & 4.4 & 1.4 & 6.2E+18 & 1.8E+21 & 8.5E-03 & 206 \\
19 & 05:13:51.33 & -67:07:42.8 & 0.27 & 3.4 & 2.1 & 5.6E+18 & 1.1E+21 & 1.2E-02 & -   \\
20 & 05:14:33.31 & -68:46:09.2 & 0.36 & 4.8 & 1.6 & 8.6E+18 & 2.8E+21 & 7.3E-03 & 213 \\
21 & 05:22:12.97 & -67:57:42.9 & 0.57 & 4.3 & 2.9 & 2.4E+19 & 3.6E+21 & 1.9E-02 & 291 \\
22 & 05:24:21.84 & -68:25:41.2 & 0.38 & 6.2 & 2.3 & 1.8E+19 & 7.0E+21 & 6.3E-03 & 350 \\
23 & 05:24:51.46 & -69:40:20.8 & 0.36 & 4.7 & 2.8 & 1.5E+19 & 4.6E+21 & 8.4E-03 & 355 \\
24 & 05:25:10.68 & -69:40:40.1 & 0.23 & 7.6 & 1.5 & 7.6E+18 & 6.5E+21 & 2.6E-03 & 358 \\
25 & 05:25:53.67 & -66:14:07.3 & 0.22 & 4.1 & 2.6 & 7.6E+18 & 3.0E+21 & 5.6E-03 & 374 \\
26 & 05:35:24.75 & -67:34:48.0 & 0.96 & 4.1 & 3.6 & 4.8E+19 & 3.6E+21 & 4.1E-02 & 451 \\
27 & 05:35:47.86 & -69:13:08.0 & 0.30 & 4.5 & 1.3 & 6.0E+18 & 3.4E+21 & 4.0E-03 & 459 \\
28 & 05:35:53.06 & -69:02:22.9 & 0.27 & 5.2 & 2.3 & 1.1E+19 & 5.3E+21 & 4.7E-03 & 462 \\
29 & 05:38:29.73 & -69:02:09.6 & 0.28 & 4.9 & 2.2 & 9.7E+18 & 4.1E+21 & 5.5E-03 & 508 \\
30 & 05:39:35.66 & -69:46:16.4 & 0.22 & 6.4 & 3.2 & 1.4E+19 & 1.0E+22 & 3.0E-03 & 531 \\
31 & 05:39:44.42 & -69:37:31.6 & 0.34 & 4.4 & 1.7 & 8.2E+18 & 3.1E+21 & 6.5E-03 & 548 \\
32 & 05:40:02.89 & -69:51:26.4 & 0.48 & 6.9 & 3.1 & 3.2E+19 & 1.2E+22 & 7.5E-03 & -   \\
33 & 05:41:16.56 & -70:55:31.9 & 0.37 & 4.8 & 2.0 & 1.3E+19 & 3.4E+21 & 9.2E-03 & 606 \\
34 & 05:43:23.42 & -69:25:12.2 & 0.20 & 5.1 & 2.0 & 6.4E+18 & 4.1E+21 & 3.5E-03 & 635 \\
35 & 05:44:42.57 & -69:28:13.6 & 0.36 & 6.0 & 2.7 & 1.9E+19 & 8.3E+21 & 5.7E-03 & 650 \\
36 & 05:46:17.45 & -69:38:26.6 & 0.27 & 3.9 & 2.3 & 7.9E+18 & 2.2E+21 & 8.1E-03 & 666 \\
37 & 05:47:01.88 & -70:46:11.0 & 0.23 & 4.1 & 3.1 & 9.0E+18 & 3.4E+21 & 5.8E-03 & 672 \\
\hline\hline
\end{tabular}
\caption{Physical parameters for the sightlines where HINSA optical depth is greater than 0.2. The Cloud ID is from Catalog C in \citet{Wong2011}. The spectra for each of these sightlines are displayed in the Appendix A.}
\label{tbl:selectedsightlines}
\end{table*}

\subsection{Spatial distribution}
\citet{Cho2016} derive a photometric metallicity map of the LMC using MCPS and OGLE III data. It shows a shallow metallicity gradient, with the central bar having the highest metallicity and the outer parts having the lowest metallicity. To confirm whether there is a corresponding trend in the spatial distribution of the HINSA-HI abundance ratio, we divide the LMC into 6 concentric elliptical rings. The radii of the rings start at 0.5 kpc, and are spaced by 0.5 kpc, doubling the bin width used by \citet{Cho2016}. The position angle (PA) of these rings is extracted from the PA measurements of \citet{Kim1998}. An HI morphologically-derived inclination angle of 22 degrees is adopted, according to the measurements of \citep{Kim1998}. The kinematic centre of the LMC HI disk is used here \citep[$05^{h}17.6^{m}, -69^{d}02^{m}$ as given by][]{Kim1998}, which deviates from the optical centre used by \citet{Cho2016} by 27 arcmin.

\begin{figure}
\centering
\includegraphics[width=3in]{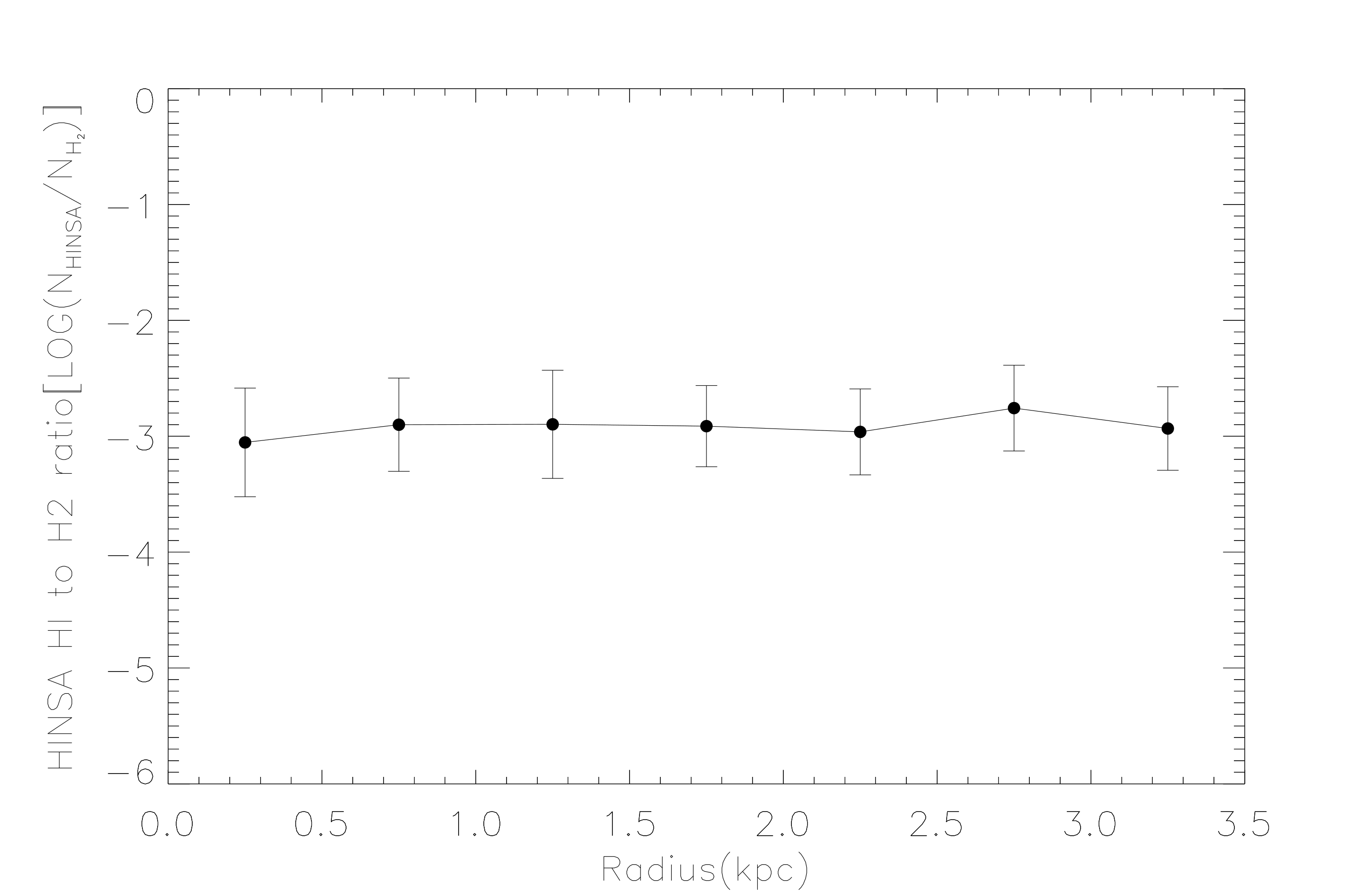}
\caption{HINSA-HI abundance as a function of the radius from the center of the LMC. The error bars show the standard deviation within in each bin.}
\label{fig:DR3HItoH2_rings}
\end{figure}

The pixels with HINSA detections are divided into 7 radial bins, containing 189, 153, 343, 442, 248, 51, 26 and 51 pixels, from small to large radius respectively. We derive the mean and standard deviation for each group by fitting the $\log_{10}$ histogram with a Gaussian. The result, shown in Figure~\ref{fig:DR3HItoH2_rings}, shows no radial gradient of the HINSA-HI abundance in the LMC.

To examine whether there is any radial trend in the HINSA-HI abundance ratio in the Milky Way, we have looked into the distances of the molecular clouds that were studied in the previous Milky Way HINSA studies \citep{LG2003,Krco2010}. We find that the existing HINSA measurements in the Milky Way are focused either on nearby molecular clouds (less than 1kpc from the Sun) or clouds at unknown distances. It is not yet possible to make a definitive statement on the radial distribution of HINSA-HI abundance in the Milky Way.

\subsection{Highlighted regions}
The detection of HINSA signatures is prevalent along the sightlines towards molecular clouds in the LMC. For those regions with strong and concentrated HINSA signatures, six have \XIIICO\ data: N11, N44, NAN17, NAN216, NAN223 and the Ridge southward of 30 Dor (the `Ridge'). The \XIIICO\ data is used to help determine the optical depth of CO, as discussed in Section 5.1. These regions are highlighted here: the optical depth map of HINSA-HI are shown in Appendix B.

The highlighted regions are mostly distributed along the two spiral features of the LMC, with one (N44) located to the north of the optical bar. The distribution of these selected regions is similar to the distribution of star formation activity in the LMC: 30 Dor and the southern ridge has the most violent star-formation activity; the western spiral feature and the region north of the optical bar are also quite active, while the region south of the optical bar is lacking major star formation activity and molecular clouds.

These maps show that the distribution of HI emission, CO and HINSA-HI roughly follows an onion shell structure with HI emission around an outer shell and HINSA-HI in the inner core. But it also seems that the spatial peak in the HINSA-HI optical depth is often mismatched with the peak of the CO cloud. This may reflect the inadequacy of CO as an \Htwo\ cloud tracer, or it may reflect an evolutionary sequence. \citet{Zuo2018} have reported the discovery of a shell structure of HINSA-HI around a molecular cloud in the Milky Way which indicates the depletion of atomic hydrogen in the center of the molecular cloud. The mismatch of the HINSA-HI peak and the CO cloud peak in LMC clouds could be due to a similar reason, although our lower spatial resolution makes this harder to judge.

\section{Discussion}
\subsection{Optical depth of CO}
In Section 3, we assumed optically thick CO emission. This assumption affects the estimate of the temperature of the HINSA gas. If the optically thick assumption breaks down, the excitation temperature of CO will be underestimated. The LTE assumption will also be incorrect so the kinetic temperature of the gas will be further underestimated.

The \XIIICO\ data for the above selected regions were therefore used to calculate the optical depth of CO to test this assumption. We selected all the pixels where the peak S/N ratio for the \XIIICO\ spectrum was larger than 3. The optical depth of \XIIICO\ for these sightlines was first calculated by assuming CO is optically thick \citep{WRH2013}:

\begin{equation}
\begin{aligned}
T_{ex}(^{12}CO)=5.5/ln(1+\frac{5.5}{T_{B}(^{12}CO)+0.82})
\end{aligned}
\label{con:Tex12CO0}
\end{equation}

Then the optical depth of \XIIICO\ is derived by

\begin{equation}
\begin{aligned}
\tau(^{13}CO)=-ln\Bigg\{1-\frac{T_{B}(^{13}CO)}{5.3}\bigg\{\\\Big[exp(\frac{5.3}{T_{ex}(^{12}CO)}-1)\Big]^{-1}-0.16\bigg\}^{-1}\Bigg\}
\end{aligned}
\label{con:tau13CO}
\end{equation}

Then the optical depth of CO is derived by multiplying the \XIIICO\ optical depth by the \XIICO/\XIIICO\ ratio:
\begin{equation}
\begin{aligned}
\tau(^{12}CO)=X(^{12}CO/^{13}CO)\tau(^{13}CO)
\end{aligned}
\label{con:tau12CO}
\end{equation}

Then a corrected excitation temperature of CO was derived by using the updated CO optical depth $\tau(^{12}CO)$:

\begin{equation}
\begin{aligned}
T_{ex}'(^{12}CO)=5.5/ln\Bigg[1+\\\Big(\frac{T_{B}(^{12}CO)}{5.5\Big[1-exp\big(-\tau(^{12}CO)\big)\Big]}+0.15\Big)^{-1}\Bigg]
\end{aligned}
\label{con:Tex12CO1}
\end{equation}

We iterated the above process (from Equation \ref{con:tau13CO} to \ref{con:Tex12CO1}) for 100 times, resulting in an improved estimate of excitation temperature and optical depth for CO.

\begin{figure}
\centering
\includegraphics[width=3.3in]{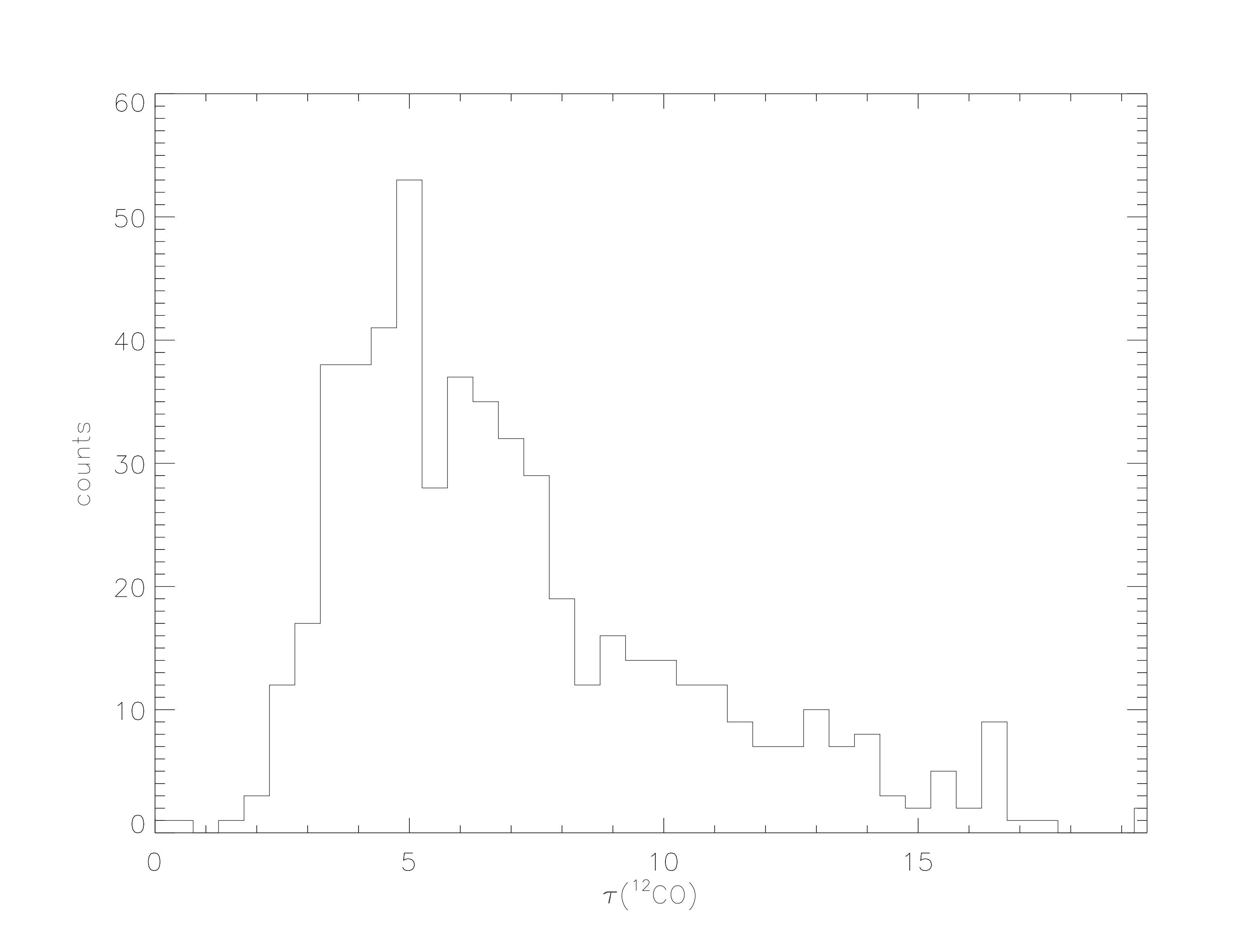}
\caption{CO optical depth distribution}
\label{fig:1213ratio}
\end{figure}

The result of the above is dependent on the value of the \XIICO/\XIIICO\ abundance ratio adopted. Previous observations have shown that the \XIICO/\XIIICO\ ratio for the molecular clouds in the LMC may or may not be different from the Milky Way value of $\sim100$. For example \citet{Joh1994} suggested a ratio of $50^{+25}_{-20}$ for N 159 in the LMC, while \cite{Isr2003} suggest a ``similar'' intrinsic isotopic ratio to the Milky Way. By adopting the conservative estimate of 50, we derived the optical depth distribution for our selected sightlines, as shown in Figure~\ref{fig:1213ratio}. The histogram peaks at $\sim5$, indicating predominantly optically thick emission. An optically thick assumption will therefore remain approximately valid for \XIICO/\XIIICO\ $> 20$. 

It is worth noting that, although the assumption that the CO is optically thick makes it straightforward to estimate \ToH, it will also result in an overestimate of CO linewidth, and an overestimate of \sigmaH. By adopting a typical set of parameters (\tauo=0.5, \sigmaH=0.5 km~s$^{-1}$, \ToH=5K) to generate an artificial HI spectrum with HINSA absorption and fitting the absorption using our method, we investigate the impact of an overestimate of \sigmaH. The simulation result suggests that when \sigmaH\ is overestimated by less than $100\%$, the corresponding peak HINSA optical depth will be underestimated by less than 50\%, but the HINSA-HI abundance will be overestimated by less than 20\%.

\subsection{Comparison with previous work}

\subsubsection{Background continuum source observations}
\citet{Dic1994}, \citet{Dic1995}, \citet{Meb1997} and \citet{MZ2000} measured 21~cm absorption lines toward 27 sources in the background of the LMC. The sparse sampling of these measurements as well as the low space filling factor of CO clouds makes it very hard for any coincidence between these two data sets: only six out of the 27 sightlines are coincident with MAGMA CO detections: 0526-678, 0536-693, 0539-696, 0540-697, 0539-697, 0521-699. Among these 6 sources, 0539-696, 0540-697, 0539-697 are located in the north part of the Ridge region, where HINSA signatures are clearly detected, whilst only 0540-697 is behind the HINSA detected pixel. The optical depth of HINSA-HI is 0.017 at this position, and \cite{Dic1994} reported the optical depth of 0540-697's four absorption components as: 1.39, 0.54, 0.57, 0.64. The HINSA feature in the sightline of 0540-697 peaks at 251 km s$^{-1}$, which is overlapping with one of the 3 subcomponents of the 237 km s$^{-1}$ component reported by \citet{Dic1994}. The clear difference in the HINSA optical depth reported here from the optical depth reported by Dickey et al. is due to the fact that we use different assumptions and thus the results trace different gaseous components in the CNM. HINSA traces the colder and thus less abundant part of the CNM. It should also be noted that, for this particular sightline, the small HINSA absorption may have a large uncertainty due to noise. It would be more meaningful if we were able to compare HINSA results from a statistical perspective.

\subsubsection{HI line modeling}
\begin{figure}
\centering
\includegraphics[width=3.3in]{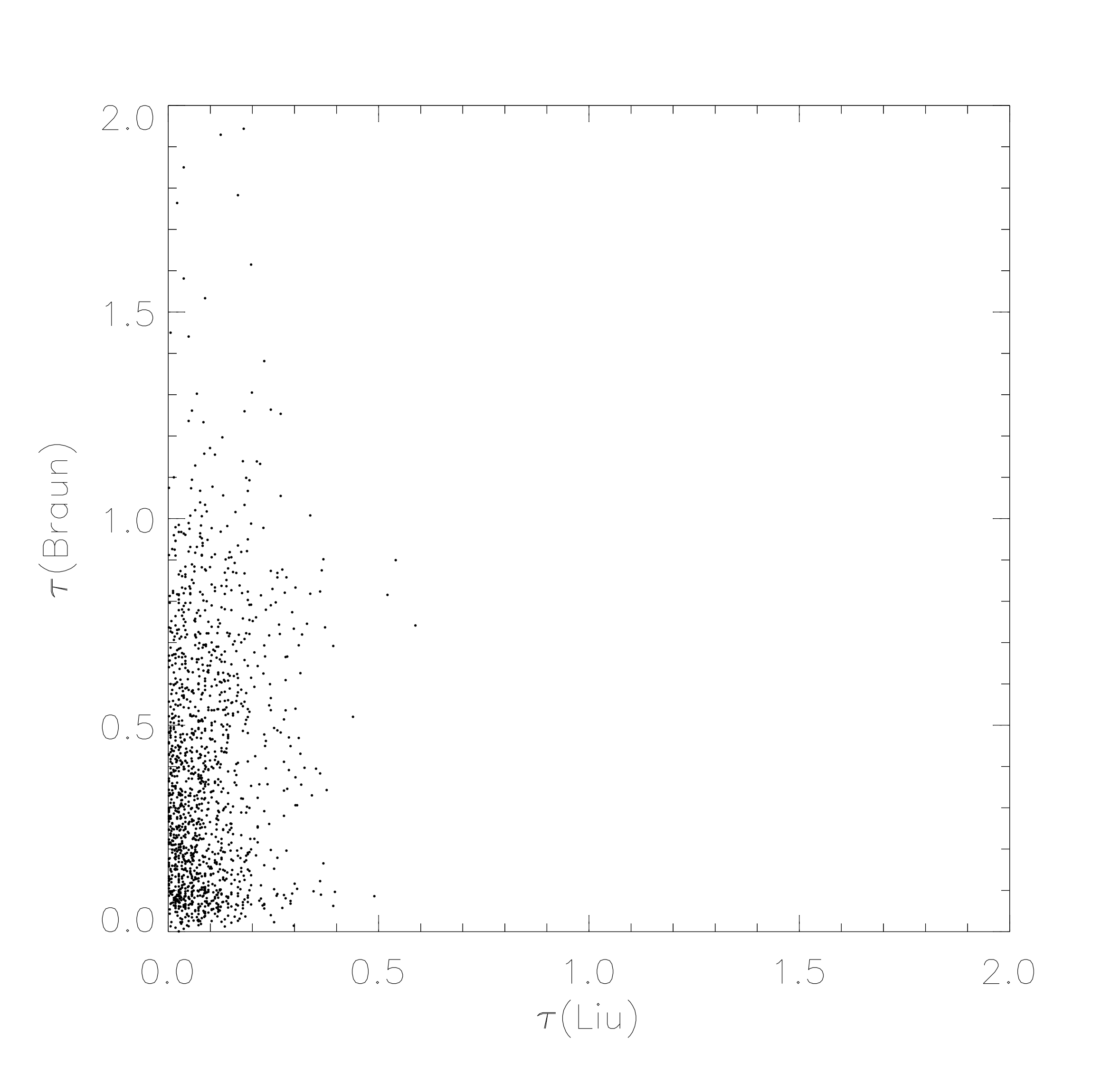}
\caption{Pixel-pixel comparison of derived optical depth value between this work and \cite{Bra2012}.}\label{fig:Tau-TauRB}
\end{figure}

\citet{Bra2012} calculated the HI optical depth of the LMC by fitting the flatness of the HI spectrum, as explained in \citet{Bra2009}. We have compared the optical depth of HINSA-HI derived in our work with the HI optical depth result of \citet{Bra2012} as shown in Figure~\ref{fig:Tau-TauRB}. No correlation is apparent. It is not surprising because the two methods are tracing different gaseous components. \cite{Bra2012} assumed the atomic clouds are isothermal on scales of 100 pc and neglected multiple velocity components, which are prevalent in the LMC. Our work, on the contrary, focuses on true optical depth effects arising from the temperature differences of molecular clouds and surrounding HI gas.

\subsubsection{Milky Way HINSA measurements}
In Section 4.1 we over-plotted the Milky Way HINSA-HI abundance result of \cite{LG2003}, \citet{Krco2010} on the histogram of the HINSA-HI abundance of the LMC. Although our result is of the same magnitude as the Milky Way results, the difference between the two data sets should be noted: the Milky Way measurements are based on data of much better spatial resolution and velocity resolution \citep[e.g. 0.13 pc and 0.16 km s$^{-1}$ for ][]{LG2003}, compared to 15 pc and 1.649 km s$^{-1}$ for the LMC.

Similar to Section 5.1, we performed a simulation to investigate the impact of HI resolution. It shows that when \ToH\ and \sigmaH\ are estimated correctly, the relatively low velocity resolution of 1.6 km s$^{-1}$ does not affect the measurement of HINSA optical depth, but it will cause a $\sim10\%$ underestimate of the HINSA-HI abundance. Since our measurements also cover a much larger volume than Galactic studies, we may also underestimate the HINSA optical depth and abundance due to the low spatial filling factor of molecular clouds. 

There are also differences in methodology in that the Milky Way studies use a Galactic rotation model to derive the dynamical distance for clouds, which can provide a relatively accurate estimate of foreground gas content. For the LMC, we can only assume the clouds are located in the middle of the warm HI disk. These differences may affect in detail the comparison of our HINSA-HI abundance results with that of the Milky Way.

It is also worth mentioning that the latest HINSA measurement by \citet{Zuo2018} has reported a relatively high HINSA-HI/\Htwo\ ratio from 0.2\% to 2\% in a single very young molecular cloud that is considered to be still in the formation process.

One of the initial assumptions of this study is that the metallicity difference between the LMC and the Milky Way may produce a measureable effect on the HINSA-HI/\Htwo\ ratio of the two galaxies. However, the insignificant difference reported in Section 4.1 does not support such a scenario. The low metallicity that results in relatively low CO abundance does not appear to significantly affect the HINSA-HI/\Htwo\ ratio. Similarly, the low metallicity that reduces the dust surface area on which \Htwo\ can form, does not affect HINSA-HI. This implies that molecular cloud cooling can still proceed despite lower dust and diffuse molecule abundance. \Htwo\ self-shielding is likely fundamental in this process.

\section{Summary}
We have used ATCA+Parkes LMC HI survey data \citep{Kim2003} and MAGMA LMC CO Survey data (DR3) \citep{Wong2011} to locate and measure HI Narrow Self-Absorption (HINSA) features towards the molecular clouds in the LMC. This is the first confirmed detection of HINSA in an external galaxy.

The HINSA-HI/\Htwo\ ratio in the LMC varies from 0.5\e{-3} to 3.4\e{-3} (68\% interval), with a mean value of $(1.31 \pm 0.03)$\e{-3}, after correcting for the effect of foreground HI gas. This is slightly higher, but not significantly different from the Milky Way value from the combined results of \citet{LG2003} and \citet{Krco2010}, namely a 68\% interval range of 0.4\e{-3} to 3.0\e{-3} and mean value of $(1.0 \pm 0.2)$\e{-3}. This result indicates similar amount of cold gas existing in the LMC compared to the Milky Way. Unlike the case for stellar metallicity, the ratio does not show a radial gradient. However, a key assumption is the accuracy of the CO X-factors that we have adopted for the Milky Way and the LMC.

The small HINSA-HI/\Htwo\ ratio shows that the molecular clouds in the LMC are more than 99 percent molecular, confirming the relatively short formation time scale of molecular clouds.

We find that HINSA features are prevalent in the surveyed sightlines: a catalog of 37 sight-lines where the peak HINSA-HI optical depth is higher than 0.2 is presented. Six typical regions where HINSA detections are concentrated (N11, N44, NAN17, NAN216, NAN223 and the LMC Ridge south of 30Dor), are examined in detail and the \XIIICO\ data for these regions are used to confirm the optical-thick assumption adopted in the calculations.

We find no correlation between our results with those based on previoursly-developed techniques, such as background continuum sources  \citep[e.g.][]{Dic1994} or HI line profile shape \citep{Bra2012}. 

\section*{Acknowledgements}
We thank the anonymous referee for useful and detailed comments. This work is supported by National Natural Science Foundation of China (NSFC) programs, No. 11988101, 11725313, 11690024, 11833008, and the CAS International Partnership Program No.114-A11KYSB20160008. The support provided by China Scholarship Council (CSC) during a visit of Boyang Liu to ICRAR/UWA is acknowledged. This work was carried out in part at the Jet Propulsion Laboratory which is operated for NASA by the California Institute of Technology. Parts of this research were supported by the Australian Research Council Centre of Excellence for All Sky Astrophysics in 3 Dimensions (ASTRO 3D), through project number CE170100013.

\bibliography{LMC2}

\appendix

\section{Spectra}

The spectra of all the sightlines listed in Table 1 are shown here. 

\begin{figure}[!htb]
\centering
\includegraphics[width=5.3in]{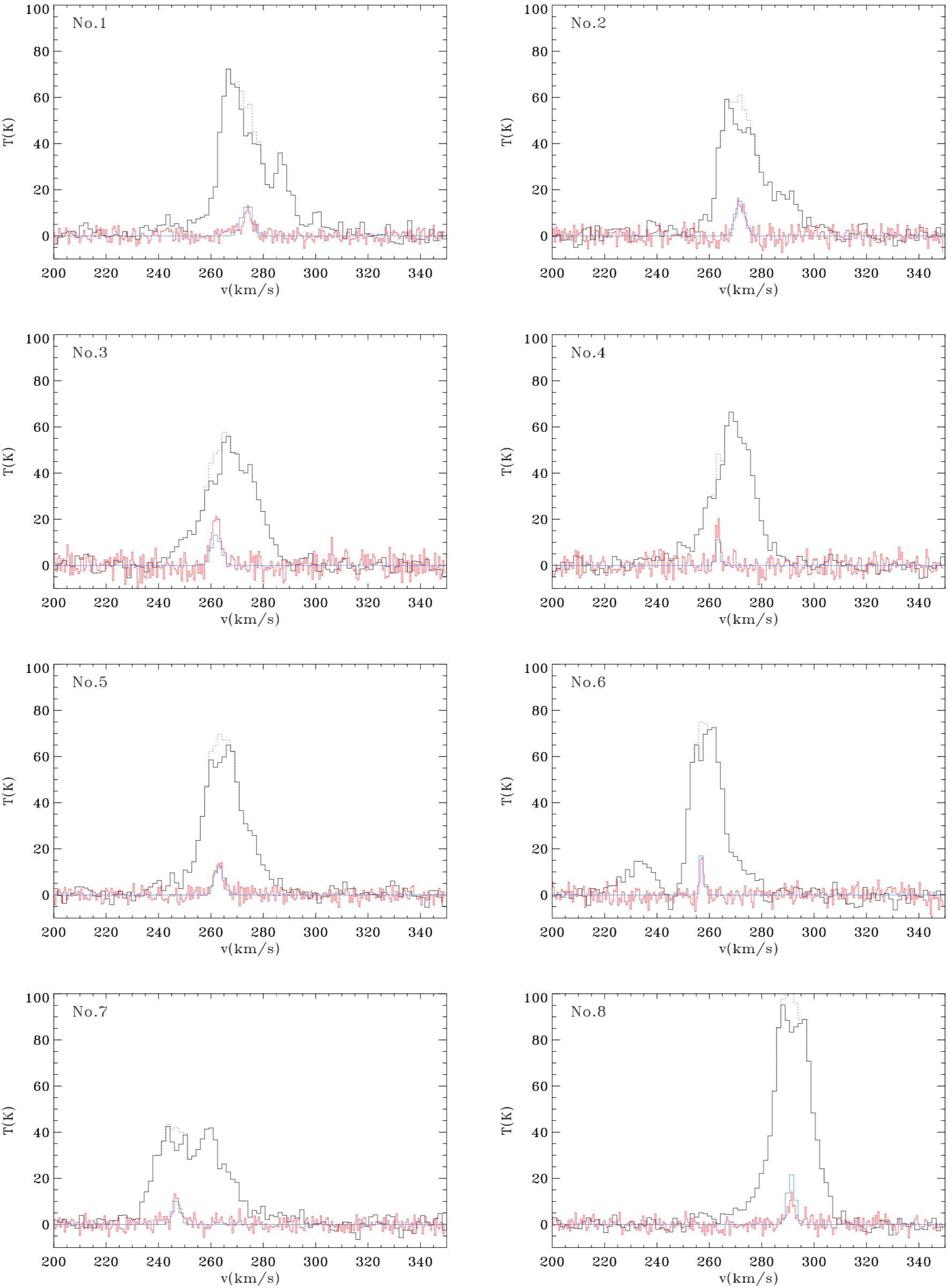}
\caption{The spectrum for each sightline listed in Table 1. The source number is labeled at the top-left corner. Solid black: HI emission spectra \citep{Kim2003};  red: CO spectra \citep{Wong2011,Wong2017}; blue: HINSA(this work); dotted black: ``original spectrum", i.e. the calculated spectrum ``before" HINSA absorption(this work, see Section 3.2.1). The vertical scale of CO spectrum is enlarged by a factor of 10. }
\label{fig:A1}
\end{figure}

\begin{figure}[!htb]
\centering
\includegraphics[width=5.5in]{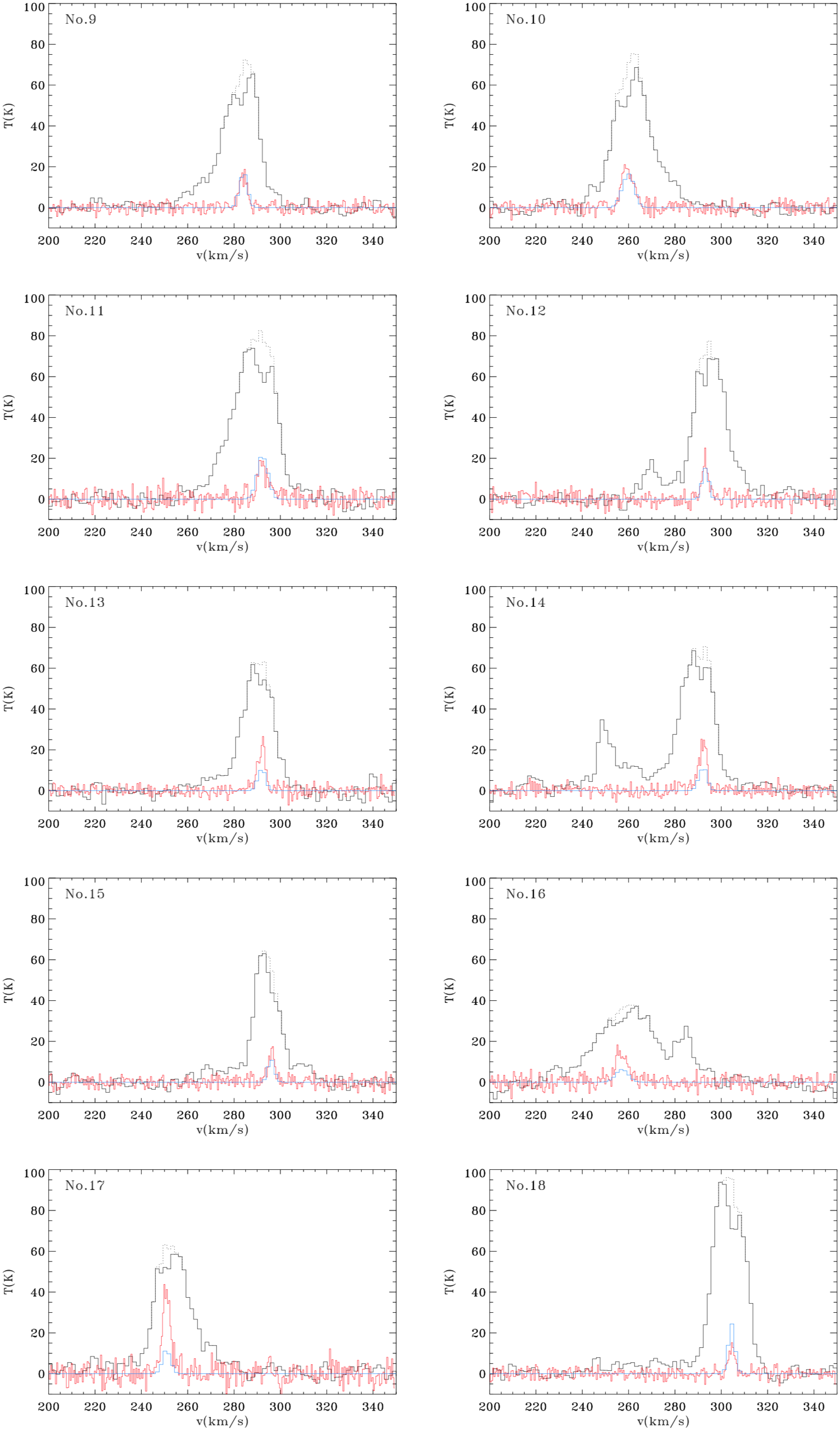}
\caption{Figure \ref{fig:A1} continued.}
\label{fig:A2}
\end{figure}

\begin{figure}[!htb]
\centering
\includegraphics[width=5.5in]{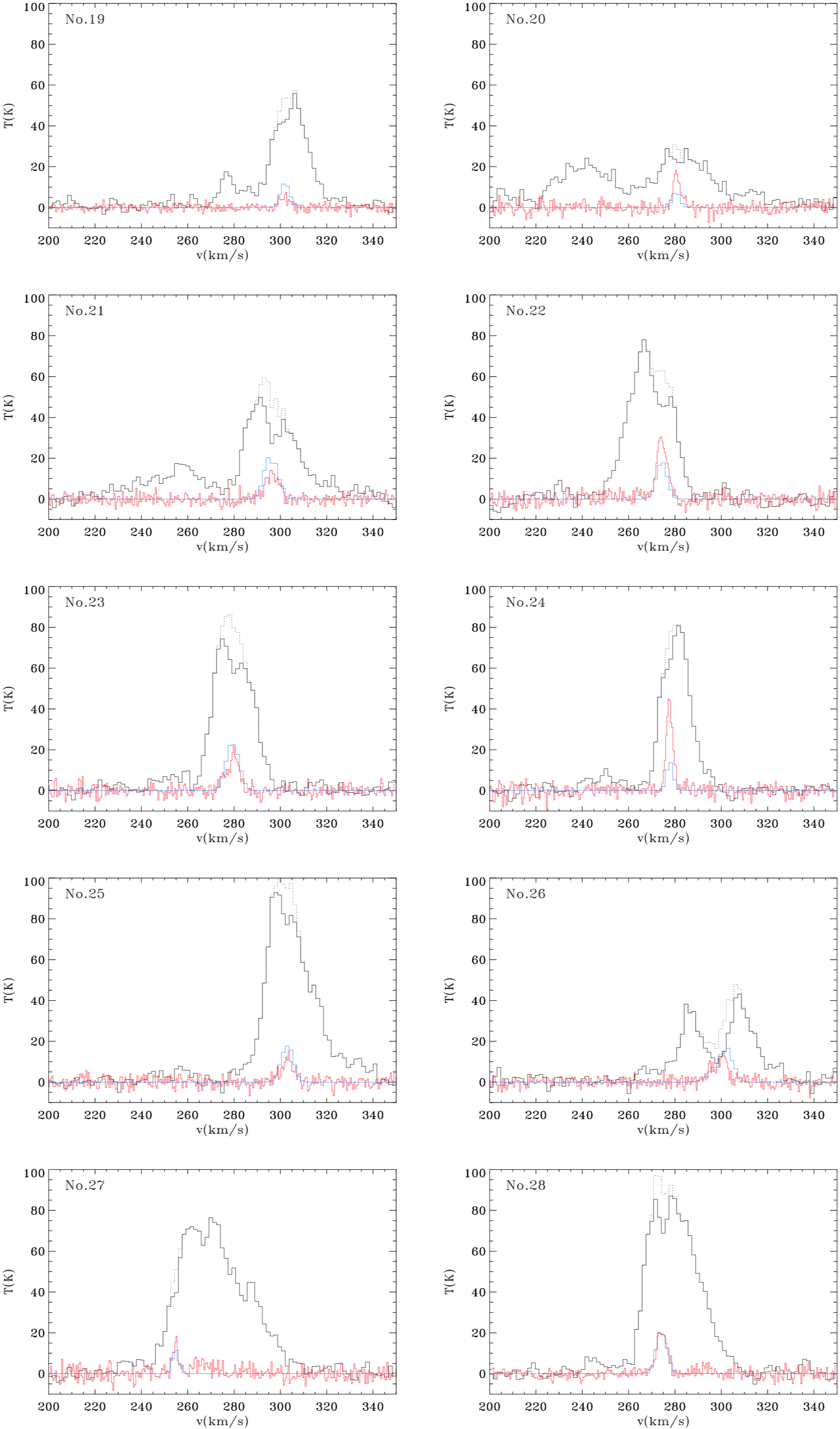}
\caption{Figure \ref{fig:A2} continued.}
\label{fig:A3}
\end{figure}

\begin{figure}[!htb]
\centering
\includegraphics[width=5.5in]{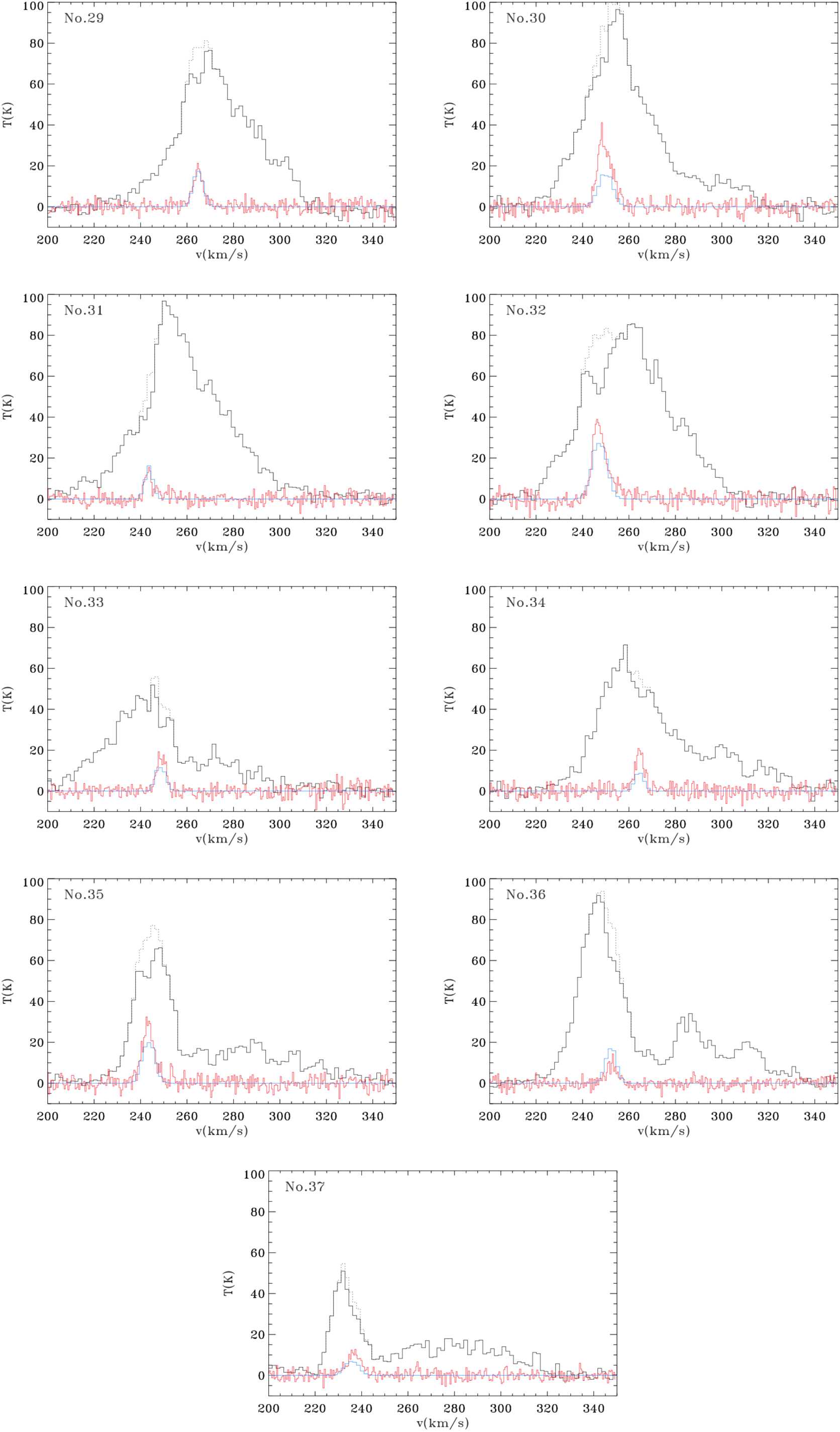}
\caption{Figure \ref{fig:A3} continued.}
\label{fig:A4}
\end{figure}

\section{Maps}

All maps of the highlighted regions (Section 4.4) are displayed here.

\begin{figure}[h]
\centering
\includegraphics[width=6.5in]{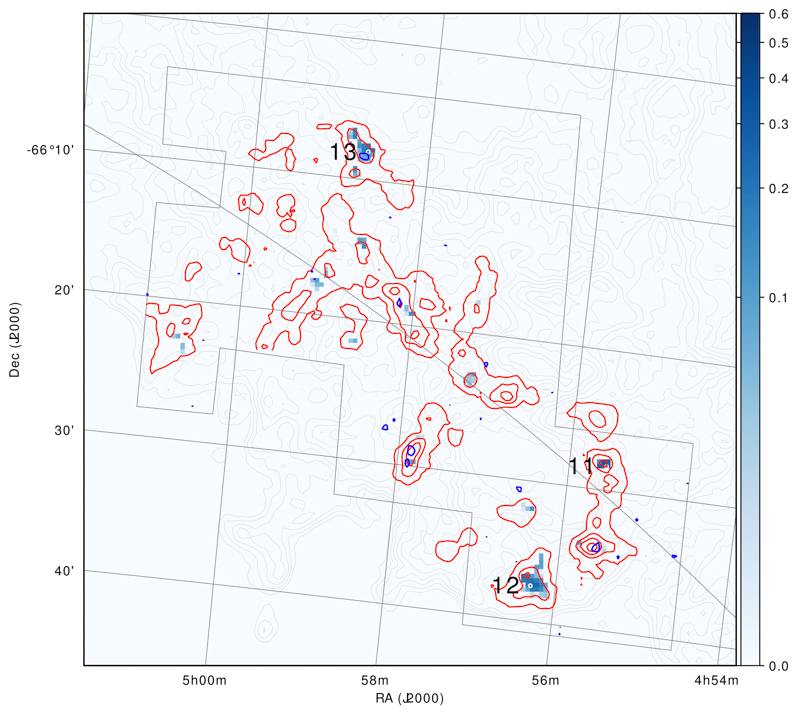}
\caption{N11 region. Blue pixels indicate the optical depth of HINSA-HI; red contours: MAGMA CO Survey DR3 moment 0 map (contour levels from 5 to 45 K$\cdot$km~s$^{-1}$); blue contours: MAGMA \XIIICO\ peak $S/N$ map for selected regions (contour levels 4.5, 6, 7.5); grey contours: peak brightness temperature of the HI emission line (contour levels from 40 to 120 K).}
\label{fig:N11}
\end{figure}

\begin{figure}
\centering
\includegraphics[width=6.5in]{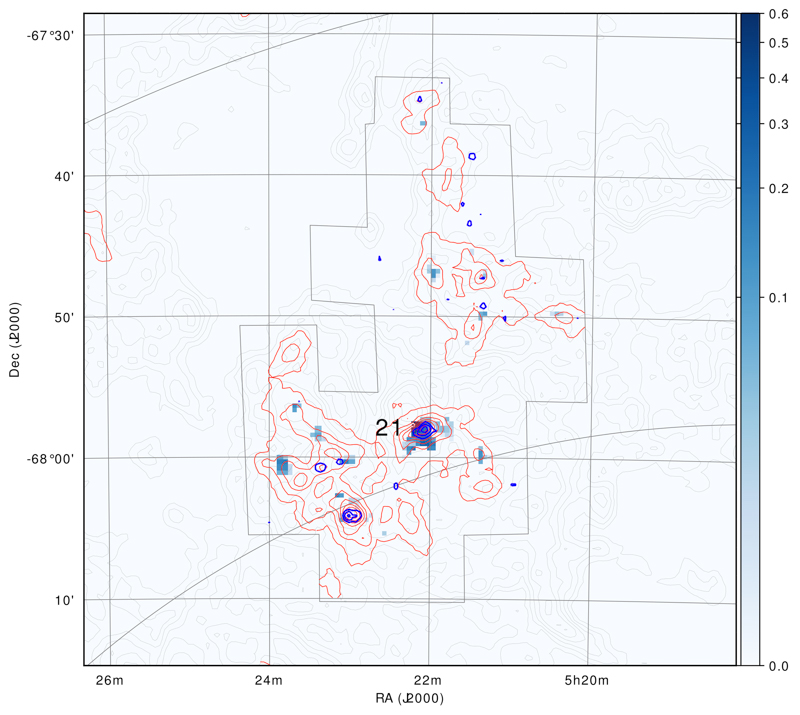}
\caption{N44 region with contours as for Figure~\ref{fig:N11}.}
\end{figure}

\begin{figure}
\centering
\includegraphics[width=6.5in]{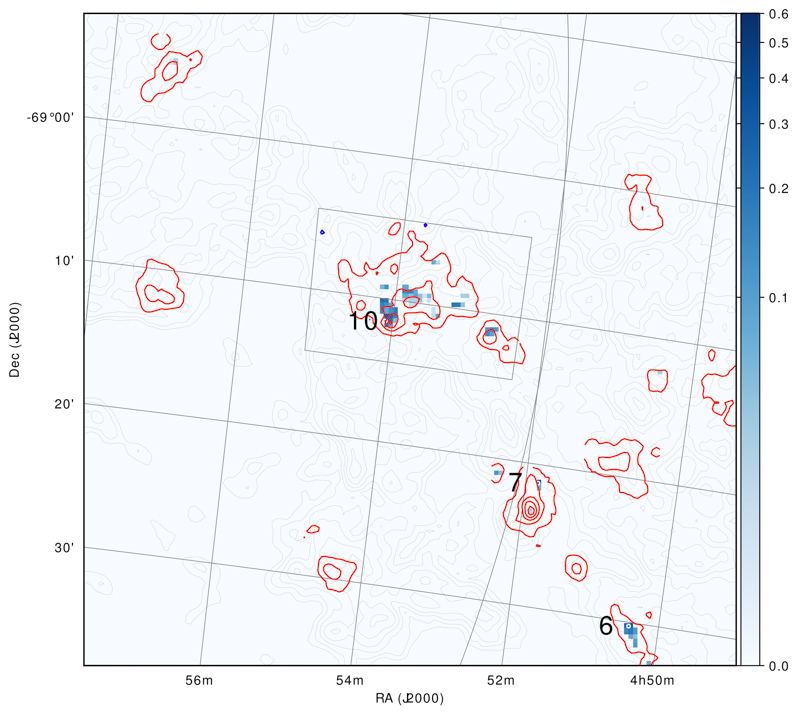}
\caption{NAN17 region with contours as for Figure~\ref{fig:N11}.}
\end{figure}

\begin{figure}
\centering
\includegraphics[width=6.5in]{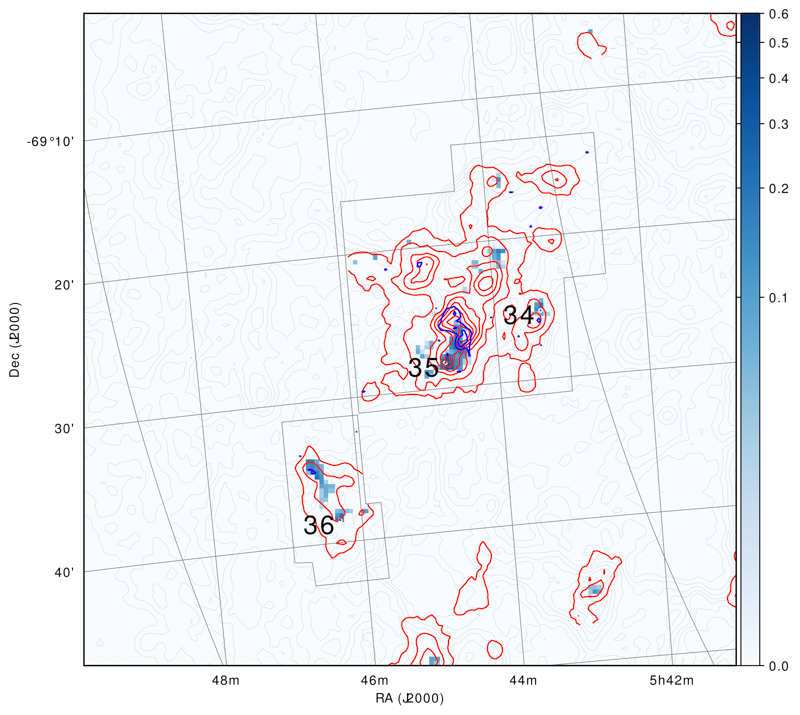}
\caption{NAN216 region with contours as for Figure~\ref{fig:N11}.}
\end{figure}

\begin{figure}
\centering
\includegraphics[width=6.5in]{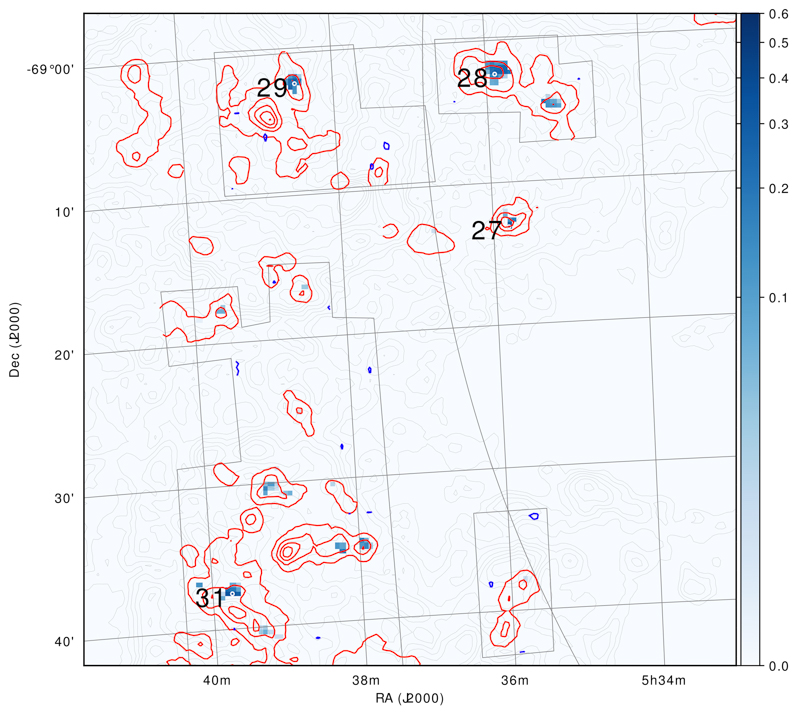}
\caption{The Ridge (north) with contours as for Figure~\ref{fig:N11}.}
\end{figure}

\begin{figure}
\centering
\includegraphics[width=6.5in]{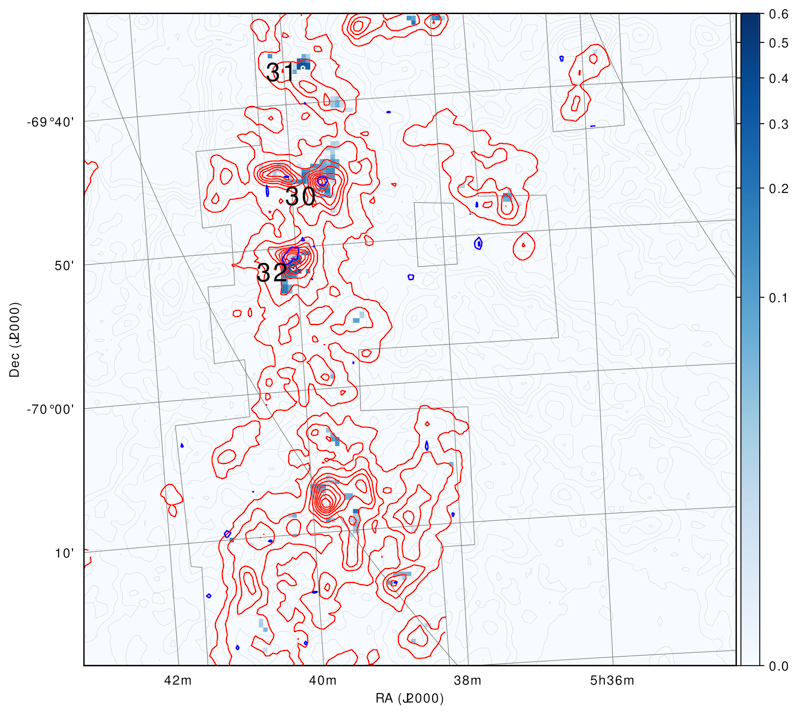}
\caption{The Ridge (middle) with contours as for Figure~\ref{fig:N11}.}
\end{figure}
\begin{figure}
\centering
\includegraphics[width=6.5in]{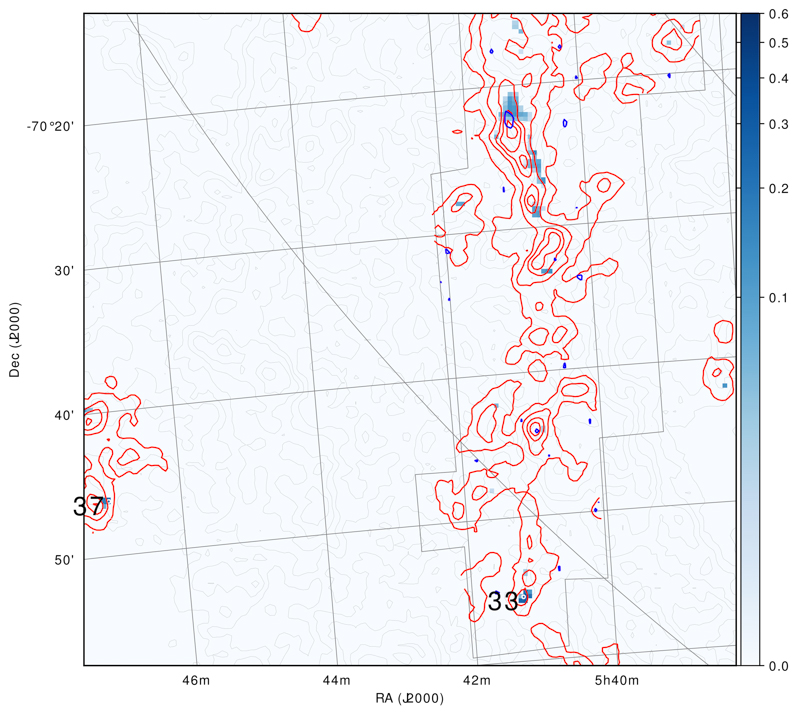}
\caption{The Ridge (south) with contours as for Figure~\ref{fig:N11}.}
\label{fig:Ridge_c}
\end{figure}

\end{document}